\documentclass[lettersize,journal]{IEEEtran}

\usepackage{amsmath,amssymb,amsfonts}
\usepackage{graphicx}
\usepackage{array}
\usepackage{booktabs}
\usepackage{xcolor}
\usepackage{url}
\usepackage[most]{tcolorbox}
\usepackage[font=footnotesize]{caption}
\usepackage[caption=false,font=footnotesize]{subfig}
\usepackage{enumitem}
\usepackage{balance}
\usepackage{algorithm}
\usepackage{algorithmic}
\usepackage{textcomp}
\usepackage{stfloats}
\usepackage{verbatim}
\usepackage{cite}
\usepackage{multirow}
\usepackage{lscape}
\usepackage{pifont}
\usepackage{cancel}
\usepackage{fontawesome5}

\usepackage{hyperref}
\hypersetup{
    colorlinks=true,
    urlcolor=blue
}

\begin{document}

\title{\textsc{PracRepair}: LLM-Empowered Automated Program Repair Inspired by Human-Like Debugging Practices}

\author{Yu~Cheng,
        Zhongxin~Liu,
        Zhenchang~Xing,
        Chao~Ni,
        Qing~Huang,
        Xiaoxue~Ren
\IEEEcompsocitemizethanks{
\IEEEcompsocthanksitem Y. Cheng, Z. Liu, C. Ni, and X. Ren are with Zhejiang University, China.
E-mail: \{yucheng1127, liu\_zx, chaoni, xxren\}@zju.edu.cn.
\IEEEcompsocthanksitem Z. Xing is with CSIRO's Data61, Australia.
E-mail: zhenchang.xing@data61.csiro.au.
\IEEEcompsocthanksitem Q. Huang is with Jiangxi Normal University, China.
E-mail: qh@jxnu.edu.cn.
\IEEEcompsocthanksitem X. Ren is the corresponding author.
}
}

\markboth{IEEE Transactions on Software Engineering,~Vol.~14, No.~8, August~2021}%
{Cheng \MakeLowercase{\textit{et al.}}: \textsc{PracRepair}: LLM-Empowered Automated Program Repair Inspired by Human-Like Debugging Practices}

\bibliographystyle{unsrt}

\maketitle
\begin{sloppypar}
\begin{abstract} 
As software systems grow in scale and complexity, debugging and repair remain costly and time-consuming. Large language models (LLMs) have advanced automated program repair (APR), but existing LLM-based APR approaches still largely rely on static or retrieved context, error messages, and coarse-grained validation outcomes. As a result, they underutilize dynamic information for failure understanding and repair, including failure-execution dynamics and patch-validation dynamics. Effectively leveraging such information, however, is challenging: failure-execution traces are large and noisy, raw static-dynamic context is not self-explanatory, and patch-validation dynamics are often reduced to coarse feedback.
To address these challenges, we propose \textsc{PracRepair}, a fully automated LLM-based APR framework inspired by human-like debugging practices. \textsc{PracRepair} constructs an on-demand static-dynamic context from buggy programs and failure executions, performs question-driven failure diagnosis to formulate explicit repair hypotheses, and iteratively refines candidate patches using validation diagnostics and trace-level behavioral changes.
Experimental results on Defects4J V1.2 and V2.0 show that \textsc{PracRepair} consistently outperforms state-of-the-art baselines. Specifically, under GPT-3.5, \textsc{PracRepair} correctly fixes 139/136 bugs on Defects4J V1.2/V2.0, while under GPT-4o it further improves to 162/171. Moreover, \textsc{PracRepair} generalizes effectively to RWB (Real-World Bugs), achieving the best performance across multiple foundation models. 
\end{abstract}


\section{Introduction}
As modern software systems continue to grow in scale and complexity, defects have become increasingly common in real-world development~\cite{Snopy, VRepair}. Fixing these defects is often challenging in practice, because in real-world software systems, the causes and effects of a defect often extend beyond a single function and require reasoning over non-local contextual information, such as call relationships, data dependencies, and execution logic~\cite{defects4j}. 

Developers typically debug in IDE-like environments~\cite{intellij_idea, vscode}, where they leverage richer information and follow a structured workflow to understand failures, formulate repair hypotheses, and iteratively refine fixes~\cite{GOULD1975151, whyline, sillito2008asking, layman2013debugging, coker2019qualitative, lou2024when}. 
More specifically, developers first gather both static and dynamic evidence by inspecting the buggy method and failing tests, navigating to relevant implementations, and tracing execution through interactive operations such as \textit{step into} and \textit{step over}. Through this process, they recover implicit execution knowledge, including call relationships, and observe fine-grained runtime behaviors such as executed paths, variable states, branch outcomes, and intermediate values~\cite{GOULD1975151, chatdbg, layman2013debugging, coker2019qualitative, lou2024when}. 
Based on such evidence, developers then diagnose failures in a question-driven manner~\cite{whyline, sillito2008asking, Alaboudi2023Hypothesizer}, asking targeted questions such as \emph{what happened here?} or \emph{why is \texttt{x} null at this point?}, and progressively narrowing down the root cause while identifying what additional evidence is needed to better understand the buggy behavior~\cite{chatdbg}. 
After completing the failure diagnosis, developers often return to the debugging environment to re-execute the patched program and compare its behavior with the original failing execution. If the patch does not fully resolve the bug, they further analyze the remaining failure and refine the repair accordingly. 
As a result, failure understanding and patch construction co-evolve through continuous feedback and refinement~\cite{chatdbg, coker2019qualitative, lou2024when}. 

Although such a debugging workflow is effective in practice, it is also expensive and time-consuming. Software developers spend roughly 35\% to 50\% of their time, and 50\% to 75\% of project budgets, on testing, verification, and debugging, costing over 100 billion dollars each year~\cite{debugging-mind-set, what-is-it-good, krasner2021cost}. 
This high cost has motivated extensive research on automated program repair (APR), which aims to automatically generate patches for buggy programs~\cite{le2019automated, Tbar, selfAPR, Tare, cure, KNOD, ITER, AlphaRepair, chatRepair, repairAgent, thinkRepair, ReInFix}. 
Early APR approaches mainly relied on manually designed fix patterns or bug-fixing datasets~\cite{Tbar, selfAPR, ITER, cure, KNOD, Tare}, but their effectiveness was often constrained by limited pattern coverage, strong data dependence, and weak generalization ability~\cite{Tbar, huang2024evolving}. Recently, large language models (LLMs) have demonstrated stronger code understanding and generation capabilities for APR~\cite{kolak2022patch, prenner2022can, AlphaRepair}. Building on this progress, recent LLM-based APR approaches, such as ChatRepair~\cite{chatRepair}, ThinkRepair~\cite{thinkRepair}, RepairAgent~\cite{repairAgent}, and ReInFix~\cite{ReInFix}, further incorporate richer repair context and iterative interaction, achieving stronger repair performance on benchmarks such as Defects4J~\cite{defects4j}. 

However, \textbf{a key limitation is that prior approaches underutilize dynamic information for failure understanding and repair, while overestimating LLMs' ability to precisely infer complex program behavior from static context alone.} 
Although recent methods such as ChatRepair~\cite{chatRepair}, ThinkRepair~\cite{thinkRepair}, RepairAgent~\cite{repairAgent}, and ReInFix~\cite{ReInFix} incorporate test feedback, richer repair context, or iterative interaction, their repair processes are still largely driven by static or retrieved context, error messages, and coarse-grained validation outcomes. 
In particular, they do not systematically exploit two types of dynamic information that are critical in practical debugging: \emph{failure-execution dynamics}, which reveal how the original failure is triggered through executed paths, runtime states, and branch outcomes, and \emph{patch-validation dynamics}, which reveal how a candidate patch changes program behavior during validation. 
Without these dynamic signals, LLMs may miss root causes or generate incomplete and overfitted fixes, especially for bugs whose root causes depend on runtime states and value evolution.

Effectively leveraging such dynamic information introduces three challenges. 
\textbf{C1: Failure-execution dynamics are large and noisy.} 
Directly exposing complete traces to the LLM may overwhelm the repair context rather than help identify failure-relevant behavior. 
\textbf{C2: Raw static-dynamic context is not self-explanatory.} 
Even when execution traces are available, the LLM still needs to determine which runtime states matter and how they relate to the faulty logic; otherwise, it may make incorrect behavioral inferences. 
\textbf{C3: Patch-validation dynamics are often underused.} 
They are frequently reduced to coarse validation outcomes, such as pass/fail results or error messages, leaving subsequent repair iterations without fine-grained evidence about what behavior has changed and why the current patch still fails.

To address the above challenges, we design and implement \textsc{PracRepair}, an LLM-empowered APR framework inspired by human-like debugging practices. 
Specifically, \textsc{PracRepair} consists of three stages. 
\textbf{(1) Static-dynamic context construction} addresses C1 by combining static program context with selectively organized execution traces collected from triggering test runs. Instead of directly exposing complete traces to the LLM, \textsc{PracRepair} indexes and structures dynamic evidence through a unified interface, allowing the LLM to access relevant code context, call relationships, executed paths, and runtime states on demand. 
\textbf{(2) Question-driven failure diagnosis} addresses C2 by guiding the LLM to ask and answer targeted diagnostic questions about what happens during execution, why the failure occurs, and how the faulty logic should be corrected. 
By retrieving the evidence needed to answer these questions, \textsc{PracRepair} progressively narrows down the root cause and formulates an explicit repair hypothesis. 
\textbf{(3) Feedback-guided patch refinement} addresses C3 by extracting validation diagnostics, code diffs, and trace diffs from failed candidate patches, and feeding these patch-validation dynamics back into diagnosis for iterative refinement. 
This enables more evidence-grounded repair and helps avoid incomplete or overfitted fixes.

Experimental results on real-world benchmarks demonstrate the effectiveness of \textsc{PracRepair}'s human-like debugging-inspired design. 
On Defects4J V1.2 and V2.0, \textsc{PracRepair} consistently outperforms state-of-the-art APR baselines~\cite{Tbar, selfAPR, Tare, KNOD, AlphaRepair, repairAgent, thinkRepair, chatRepair, ReInFix}. Specifically, under GPT-3.5, \textsc{PracRepair} correctly fixes 139 bugs on Defects4J V1.2 and 136 bugs on Defects4J V2.0; under GPT-4o, it further improves to 162 and 171 bugs, respectively. 
It also fixes many previously unsolved bugs, including 75 unique correct fixes achieved by \textsc{PracRepair} with GPT-3.5 and 93 unique correct fixes under GPT-4o when compared with ReInFix. 
Across repair scenarios, \textsc{PracRepair} performs strongly from single-line to multi-function bugs, with particularly notable advantages on more challenging cases. Ablation studies further verify the effectiveness of all three stages, showing that these gains come from enriching repair with failure-aware information, question-driven diagnosis, and feedback-guided refinement inspired by human-like debugging practices. Beyond Defects4J, \textsc{PracRepair} also generalizes well to RWB V1.0/V2.0~\cite{thinkRepair}, achieving the best performance across multiple foundation models.

Overall, this work makes the following main contributions: 

\enlargethispage{\baselineskip}
\begin{itemize}
\item
We identify and formulate a critical gap between practical debugging workflows and existing LLM-based APR techniques. Specifically, we show that current approaches have not fully exploited three debugging practices that are widely used by developers: static-dynamic evidence gathering, question-driven failure diagnosis, and feedback-guided patch refinement. 

\item 
We introduce \textsc{PracRepair}, a fully automated LLM-based APR framework that operationalizes these debugging practices. \textsc{PracRepair} constructs an on-demand static-dynamic context from buggy programs and failure executions, performs question-driven failure diagnosis to formulate explicit repair hypotheses, and iteratively refines candidate patches using validation diagnostics and trace-level behavioral changes. 

\item 
We conduct extensive experiments on Defects4J and RWB. Results show that \textsc{PracRepair} consistently outperforms SOTA APR baselines, remains effective across different repair scenarios from single-line to multi-function bugs, and generalizes well across multiple foundation models. Ablation studies further confirm the effectiveness of all three modules.

\end{itemize}

The code and experimental results are available at \href{https://doi.org/10.5281/zenodo.19336422}{\textcolor{blue}{[Link]}}
\section{Motivation}

\begin{figure*}[t]
    \centering
    \includegraphics[width=0.9\textwidth]{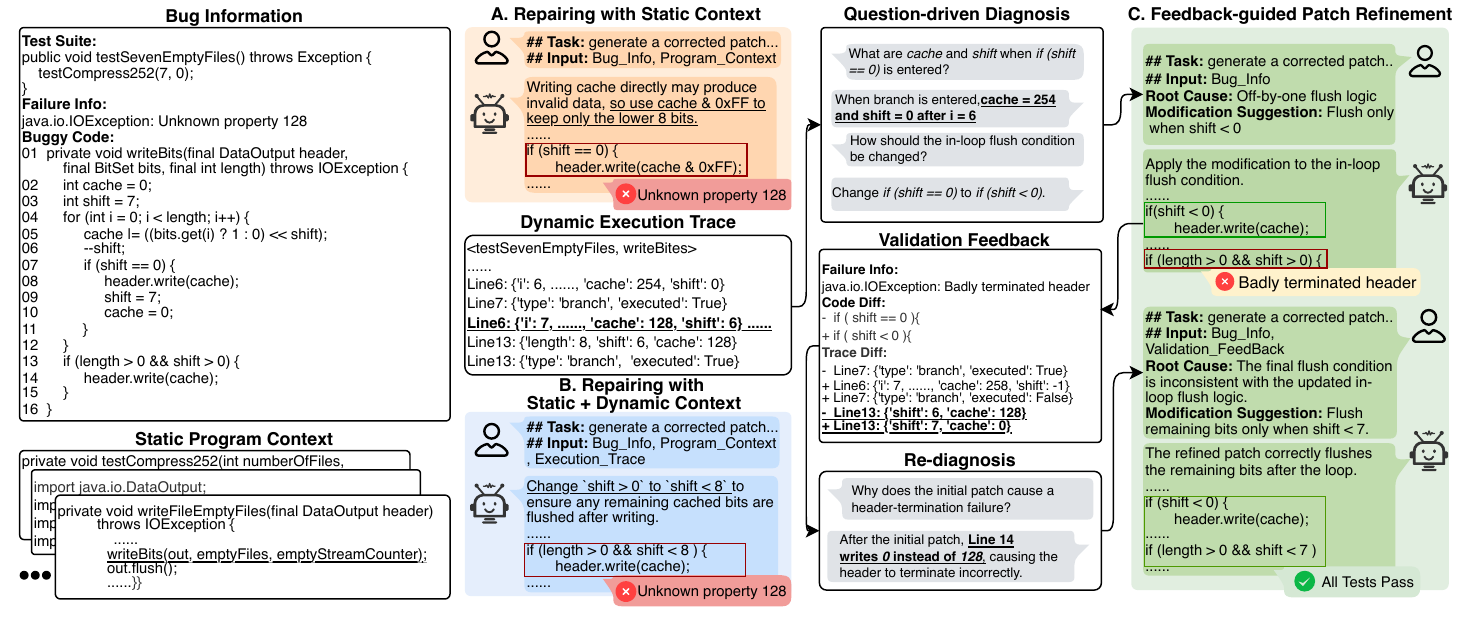}

    \centering
    \caption{A motivating example based on Compress-21, illustrating the need for dynamic execution trace, question-driven diagnosis, and feedback-guided refinement in APR.}
    \label{fig:motivation_example}
\end{figure*}

We present a real-world example in Figure~\ref{fig:motivation_example}, based on a simplified code snippet excerpted from the Java project \href{https://commons.apache.org/proper/commons-compress/}{\emph{commons-compress}}, to illustrate why APR should move beyond direct patch generation and instead follow a debugging-oriented repair process. 
This example mirrors how developers debug in practice: they inspect static code context, observe concrete runtime states, ask targeted diagnostic questions, and refine incomplete fixes based on changed program behavior. 
Accordingly, it motivates the three key designs of \textsc{PracRepair}: static-dynamic context construction, question-driven failure diagnosis, and feedback-guided patch refinement.
The bug is located in the \texttt{writeBits} method, which packs bits into a temporary buffer \texttt{cache} and flushes them according to \texttt{shift}. 
The defect is an off-by-one error: \texttt{shift} is initialized to 7 and decremented after each bit is written, but the buggy code flushes the buffer when \texttt{shift == 0}, causing only seven bits to be written and triggering \texttt{java.io.IOException: Unknown property 128}.

\noindent
\textbf{C1: Failure-execution dynamics are large and noisy.} 
Existing LLM-based APR methods already use failure-related signals, but these signals are often coarse-grained. 
For example, ChatRepair~\cite{chatRepair} and ThinkRepair~\cite{thinkRepair} use buggy code, failing tests, and validation feedback, but do not expose fine-grained execution traces such as executed paths, variable states, and branch outcomes. 
As shown in the \emph{A. Repairing with Static Context} panel, given only bug information and static context, the model changes \texttt{header.write(cache)} to \texttt{header.write(cache \& 0xFF)}. 
This patch appears to address the symptom suggested by \texttt{Unknown property 128}, but still fails with the same error. 
In contrast, the \emph{Dynamic Execution Trace} panel reveals how \texttt{cache} and \texttt{shift} evolve across loop iterations and exposes the runtime state where the failure is triggered. 
However, complete execution traces in real programs may contain many irrelevant calls, branches, and state changes, and directly exposing them to the LLM may overwhelm the repair context. 
This motivates \emph{static-dynamic context construction} in Stage I of \textsc{PracRepair}. 

\noindent
\textbf{C2: Raw static-dynamic context is not self-explanatory.}
Recent agentic APR methods, such as RepairAgent~\cite{repairAgent} and ReInFix~\cite{ReInFix}, allow the model to interact with external tools, retrieve additional context, or refine patches iteratively. 
However, richer context alone does not guarantee that the model will identify the failure-relevant behavior. 
To illustrate this issue, the \emph{B. Repairing with Static + Dynamic Context} panel shows what may happen when the model is provided with additional dynamic evidence without explicit diagnostic guidance. 
This suggests that raw context is useful but not self-explanatory: the model still needs to determine which runtime states matter and how they explain the faulty logic. 
In the \emph{Question-driven Diagnosis} panel, targeted questions such as \emph{What are \texttt{cache} and \texttt{shift} when \texttt{if (shift == 0)} is entered?} and \emph{How should the in-loop flush condition be changed?} guide the model to focus on the premature flush and formulate a more precise repair hypothesis. 
This motivates \emph{question-driven failure diagnosis} in Stage II of \textsc{PracRepair}.

\noindent
\textbf{C3: Patch-validation dynamics are often underused.}
Iterative APR methods commonly use validation results to refine patches~\cite{chatRepair, thinkRepair, repairAgent}, but validation feedback is often reduced to coarse outcomes such as pass/fail results or error messages. 
As shown in the \emph{Validation Feedback} panel, the initial patch changes \texttt{if (shift == 0)} to \texttt{if (shift < 0)}, which removes the original failure \texttt{Unknown property 128} but introduces a new failure, \texttt{Badly terminated header}. 
If validation is treated only as a pass/fail signal, the model receives limited guidance for the next repair attempt. 
Instead, \textsc{PracRepair} extracts structured validation feedback, including the validation diagnostic, code diff, and trace diff between the original and patched executions. 
The trace diff reveals that after the initial patch, the post-loop write becomes inconsistent with the updated in-loop flush behavior, localizing the remaining issue to the final flush condition. 
This leads to the refined patch \texttt{if (length > 0 \&\& shift < 7)}, which passes all tests. 
This motivates \emph{feedback-guided patch refinement} in Stage III of \textsc{PracRepair}. 
\section{Approach}

\begin{figure*}[t]
    \centering
    \includegraphics[width=0.9\textwidth]{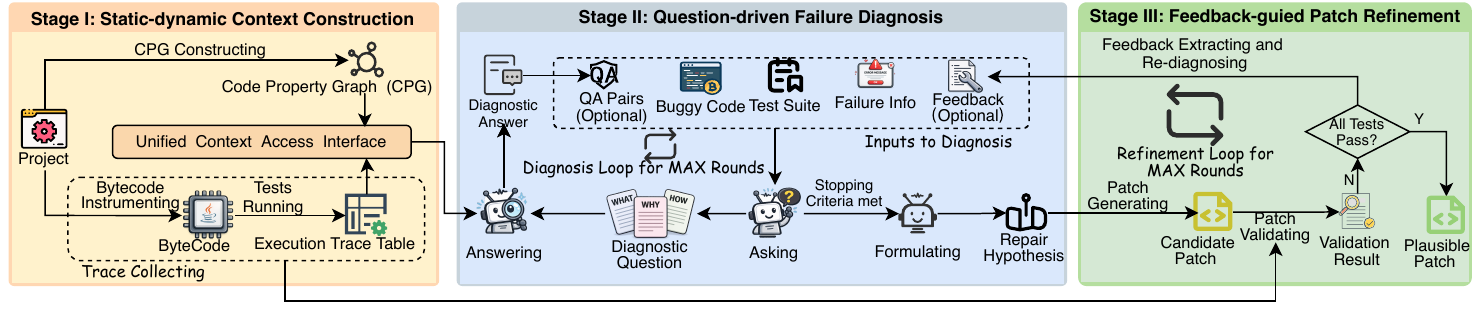}
    \captionsetup{justification=centering,singlelinecheck=false}
    \caption{The Overall Framework of \textsc{PracRepair}.}
    \label{fig:approach}
\end{figure*}

Figure~\ref{fig:approach} shows the overall workflow of \textsc{PracRepair}, which aims to improve automated program repair by drawing inspiration from human-like debugging practices. 
To achieve this goal, \textsc{PracRepair} first extracts static program context from the project and collects dynamic execution trace from triggering test runs to build a context basis for repair, 
while providing a uniform interface for the LLM to access the needed information on demand; instead of directly using this evidence for patch generation, 
it then guides the LLM to diagnose faulty program behaviors by incrementally raising and answering diagnostic questions, thereby formulating an explicit repair hypothesis; 
finally, it generates and validates candidate patches, analyzes the code-level and behavioral differences introduced by each patch, and feeds these diagnostic signals back into failure diagnosis to iteratively refine the repair. 
Specifically, \textsc{PracRepair} contains three main stages: \emph{Static-dynamic Context Construction} (Stage I), \emph{Question-driven Failure Diagnosis} (Stage II), and \emph{Feedback-guided Patch Refinement} (Stage III). 
During repair, \textsc{PracRepair} maintains three intermediate artifacts: the diagnostic QA history, the repair hypothesis, and the validation feedback. 
The diagnosis loop updates the QA history by asking and answering one diagnostic question at a time, and terminates when no further question is needed or the diagnosis budget is exhausted. 
The refinement loop updates the repair hypothesis using feedback from failed candidate patches, and terminates when a plausible patch is found or the refinement budget is exhausted. 
In our implementation, the diagnosis and refinement budgets are set to 10 and 3, respectively, and both serve as upper bounds rather than mandatory numbers of rounds.

\subsection{Static-dynamic Context Construction}
\label{sec:information_integration}

As shown in Figure~\ref{fig:approach}, the goal of \emph{Static-dynamic Context Construction} is to build a failure-relevant context basis for subsequent diagnosis and repair. To simulate how developers debug in practice, \textsc{PracRepair} must support the LLM in understanding both \emph{where} failure-relevant logic resides in the program and \emph{how} the faulty behavior is actually triggered during execution. This requires two complementary sources of evidence. Static information is needed to expose the structural and semantic context of the bug, such as surrounding implementations, control structures, call relationships, and value-flow dependencies. Dynamic information is needed to reveal the concrete failure behavior at runtime, including executed paths, branch outcomes, variable states, and failure-triggering execution conditions. To unify these two complementary sources for diagnosis and repair, \textsc{PracRepair} provides a uniform interface to access the required information. 

\subsubsection{Static context construction via CPG construction.}
To support diagnosis of failure-relevant program structure and semantics, \textsc{PracRepair} first performs static program analysis on the input project and constructs a Code Property Graph (CPG). We use \emph{Joern}~\cite{joern} to parse the project source code and build the CPG~\cite{yamaguchi2014cpg}, which unifies the abstract syntax tree (AST), control-flow graph (CFG), and data-dependence relations into a single representation. Based on this representation, \textsc{PracRepair} can access not only syntactic entities such as classes, methods, and statements, but also semantic relations such as control branches, call edges, and variable definition--use chains. This static evidence is important for understanding the structural context of the buggy code, locating related program entities, tracing inter-procedural dependencies, and reasoning about how values and control decisions propagate to failure-relevant locations.

\subsubsection{Dynamic context construction via trace collection.}
To support the diagnosis of faulty behavior, \textsc{PracRepair} further collects runtime execution evidence from triggering test executions. Since the goal is to observe actual failing behavior without modifying source code semantics, we adopt non-intrusive bytecode instrumentation~\cite{Khatchadourian2019safe}. 
Specifically, \textsc{PracRepair} instruments Java bytecode using JavaAgent~\cite{javaagent_instrumentation} and ASM~\cite{lenglet2002asm}, and then executes the triggering tests to record runtime states. Considering that dynamic execution information can be extremely large in real-world programs, \textsc{PracRepair} focuses trace collection on the buggy function under triggering test executions, so as to capture failure-relevant runtime behavior while controlling trace noise and token overhead. 
We apply statement-level instrumentation to capture execution evidence with sufficient granularity for diagnosis while controlling trace noise. \textsc{PracRepair} records the executed statement sequence within the buggy function, the values of in-scope variables after each executed statement, and the outcomes of conditional branches. 
For object-type variables, fields are recursively serialized up to a depth of 3 to balance contextual richness and token efficiency. 
The collected runtime evidence is organized into an \emph{Execution Trace Table}, where each table corresponds to a specific \textless triggering test, buggy function\textgreater{} pair and records the executed statements, their associated runtime states, and branch outcomes.

\begin{table*}[t]
\centering
\setlength{\tabcolsep}{4pt}
\small
\renewcommand{\arraystretch}{1.15}
\caption{Unified interface for accessing static and dynamic context in \textsc{PracRepair}}
\label{tab:function_calls}
\resizebox{1\textwidth}{!}{
\begin{tabular}{lll}
\toprule
\textbf{Access Capability} & \textbf{Function Calls} & \textbf{Diagnostic Use} \\
\midrule

Dependency \& entity localization 
& \texttt{get\_imports\_of\_path}; \newline \texttt{find\_class}; \newline \texttt{find\_method} 
& Support \emph{why}- and \emph{how}-type questions by locating referenced entities and dependencies \\

Structured definitions \& code inspection 
& \texttt{get\_definition\_of\_class}; \newline \texttt{get\_definition\_of\_method}; \newline \texttt{get\_code\_of\_method} 
& Support \emph{how}-type questions by inspecting surrounding logic and candidate modification points \\

Structural \& semantic relation inspection 
& \texttt{get\_structure\_of\_method}; \newline \texttt{get\_callers\_of\_method}; \newline \texttt{get\_def\_use\_of\_variable} 
& Support \emph{why}-type questions by analyzing control flow, caller relations, and value propagation \\

Execution-path inspection 
& \texttt{get\_execution\_path} 
& Support \emph{what}-type questions by revealing what actually happens during failing execution \\

Runtime-value inspection 
& \texttt{get\_runtime\_values} 
& Support \emph{what}- and \emph{why}-type questions by tracking variable evolution and abnormal states \\

Statement-level state inspection 
& \texttt{get\_state\_at\_statement} 
& Support \emph{what}-, \emph{why}-, and \emph{how}-type questions by examining concrete states and checking repair hypotheses \\

\bottomrule
\end{tabular}
}
\end{table*}

\subsubsection{Unified context access interface.}
The static and dynamic context constructed above is not provided to the LLM all at once. Instead, \textsc{PracRepair} exposes it through a uniform interface that supports on-demand retrieval during diagnosis. This design avoids overwhelming the LLM with the full project context and long execution traces, while allowing context retrieval to be guided by the current diagnostic need, similar to how developers inspect code and execution behavior during debugging. 

For static evidence, the interface supports three forms of access: \textbf{(1) dependency and entity localization}, which helps identify relevant program entities and resolve referenced types; \textbf{(2) structured definitions and code inspection}, which helps inspect classes, methods, and implementations to understand surrounding logic and identify candidate modification points; and \textbf{(3) structural and semantic relation inspection}, which helps reason about control constructs, caller relationships, and variable definition--use chains. For dynamic evidence, the interface supports three common debugging needs: \textbf{(1) execution-path inspection}, which helps understand what actually happens during failing execution; \textbf{(2) runtime-value inspection}, which helps track variable evolution and identify abnormal state changes; and \textbf{(3) statement-level state inspection}, which helps examine concrete program states at specific locations and check whether a repair hypothesis is consistent with the observed execution.

Table~\ref{tab:function_calls} summarizes the function calls that implement these context access capabilities. Together, the static and dynamic evidence constructed in this stage, along with the uniform access interface, provide the information basis for the next stage, \emph{Question-driven Failure Diagnosis} (Section~\ref{sec:question-driven_failure_diagnosis}).

\subsection{Question-driven Failure Diagnosis}
\label{sec:question-driven_failure_diagnosis}
As illustrated in Stage II of Figure~\ref{fig:approach}, \textsc{PracRepair} does not directly generate a patch from the context constructed in Section~\ref{sec:information_integration}. 
Instead, it first transforms the collected evidence into diagnostic understanding through question-driven failure diagnosis. 
This stage takes as input the buggy code, test suite, failure information, the accumulated diagnostic QA history, and optionally the validation feedback returned from Stage III. 
At each round, \textsc{PracRepair} raises one diagnostic question, retrieves the evidence needed to answer it, and appends the resulting QA pair to the diagnosis history. 
Each QA pair records the question, retrieved evidence, diagnostic answer, and repair implication. 
The loop terminates when the diagnosis budget is exhausted or the accumulated QA history is sufficient to formulate a repair hypothesis. 
Finally, \textsc{PracRepair} summarizes the diagnostic findings into an explicit repair hypothesis.

\subsubsection{Question Asking.}
Prior studies show that questions are central to debugging and program understanding~\cite{whyline, sillito2008asking}. Accordingly, \textsc{PracRepair} reduces diagnostic uncertainty through three question types, grounded in the context access capabilities in Table~\ref{tab:function_calls}.

\begin{itemize}
\item \textbf{What-type questions} establish factual understanding of the failing execution, such as executed statements, branch outcomes, variable evolution, and deviations from expected behavior, mainly using dynamic evidence.

\item \textbf{Why-type questions} explain the failure by connecting abnormal runtime behavior to underlying program logic, such as incorrect control flow, abnormal state transitions, or invalid data dependencies, using both static and dynamic evidence.

\item \textbf{How-type questions} determine how to change the faulty logic to restore the intended semantics, mainly using the diagnosed root cause and static code context.
\end{itemize}

To avoid unnecessary diagnostic overhead, \textsc{PracRepair} does not ask all questions at once. 
Instead, it requires the LLM to make a structured diagnostic decision at each step. 
The decision either raises one new diagnostic question or returns a stopping signal. 
When raising a question, the LLM must specify the question type, the target program entity or runtime behavior to inspect, and the evidence needed to answer the question. 
For example, a what-type question may target the value of a variable at a suspicious statement, while a why-type question may target the control or data dependency that explains an abnormal state. 
This constrained format makes the diagnosis process traceable and prevents the model from asking multiple unrelated questions in one round. 
The diagnosis loop terminates when either the diagnosis budget is reached or the LLM returns the stopping signal.

\subsubsection{Question Answering.}
To answer each diagnostic question, \textsc{PracRepair} lets the LLM retrieve failure-relevant context through the unified interface in Section~\ref{sec:information_integration}. 
Depending on the question type, the LLM may inspect static evidence, such as dependencies, implementations, and structural relations, or dynamic evidence, such as execution paths, runtime values, and statement-level states. 
Inspired by ReAct~\cite{yao2023react}, this process interleaves reasoning and retrieval until enough evidence is collected to answer the question. 
The resulting answer is paired with the question and appended to the QA history for subsequent diagnosis. 
When Stage III returns validation feedback, the same loop incorporates it to re-diagnose the current patch behavior.

\subsubsection{Repair Hypothesis Formulating.}
When the diagnosis loop terminates, \textsc{PracRepair} formulates an explicit \emph{repair hypothesis} based on the accumulated QA history and the currently available failure-relevant evidence. 
The hypothesis is represented in a structured form with four fields: \emph{faulty behavior}, which describes the observed abnormal execution; \emph{supporting evidence}, which records the key QA findings and retrieved context; \emph{suspected root cause}, which explains why the failure occurs; and \emph{modification suggestion}, which specifies how the faulty logic should be changed. 
For example, in Figure~\ref{fig:motivation_example}, the hypothesis identifies premature flushing at \texttt{shift == 0} as the faulty behavior, uses the observed values of \texttt{cache} and \texttt{shift} as supporting evidence, and suggests changing the in-loop flush condition. 
This structured hypothesis serves as the output of Stage II and the input to Stage III, bridging diagnosis and patch generation.

\subsection{Feedback-guided Patch Refinement}
As illustrated in Stage III of Figure~\ref{fig:approach}, \textsc{PracRepair} turns the \emph{repair hypothesis} produced by \emph{Question-driven Failure Diagnosis} into an iterative loop. 
Rather than treating validation as a simple pass/fail check, this stage explicitly analyzes the behavioral differences before and after patching and uses them as new evidence for subsequent diagnosis. 
In this way, Stage III closes the loop between repair and diagnosis: a repair hypothesis guides patch generation, patch validation reveals how the patched execution differs from the original failing execution, and unsuccessful validation produces feedback that is fed back into Stage II to refine the diagnosis and the next repair hypothesis. 
This refinement loop continues until the maximum number of refinement rounds is reached (i.e., 3), or terminates earlier once a plausible patch is found. 

\subsubsection{Patch Generating.}
Given the current \emph{repair hypothesis}, \textsc{PracRepair} prompts the LLM to generate a candidate patch for the buggy function. The patch-generation prompt is constructed from three parts: \textbf{(1) bug context}, including the buggy function, triggering tests, and failure information; \textbf{(2) repair hypothesis}, including the suspected root cause of the failure and the corresponding modification suggestions; and \textbf{(3) generation instruction}, which directs the LLM to produce a corrected implementation of the buggy function. In this way, patch generation is guided not only by the observed symptom, but also by the explicit diagnostic understanding accumulated in Stage II. To preserve input clarity and minimize prompt bias, \textsc{PracRepair} adopts a zero-shot prompting strategy, with patch generation relying solely on the structured prompt rather than in-context examples. Due to page limits, all AI prompt templates are provided in the artifact~\cite{pracrepair_artifact}.

\subsubsection{Patch Validating.}
After generating a candidate patch, \textsc{PracRepair} applies it to the original program and validates the patched program through compilation and test execution. During this process, \textsc{PracRepair} also collects execution traces from the patched program using the same trace collection procedure described in Section~\ref{sec:information_integration}, so that patched behaviors can later be compared with the original failing execution. If the patched program compiles successfully and passes all tests within the maximum execution time (i.e., 10 minutes), the patch is regarded as a \emph{plausible patch}.

\subsubsection{Feedback Extracting and Re-diagnosing.}
If a candidate patch does not pass validation, \textsc{PracRepair} does not treat the result as a simple failure signal. Instead, it first determines \emph{how} the current repair attempt fails, because different validation outcomes provide different high-level directions for the next diagnosis round. For example, a compilation failure indicates that the patch itself is syntactically or semantically invalid. 

To provide such high-level guidance, \textsc{PracRepair} first categorizes invalid validation results into four outcomes: \textbf{(1) compilation failures}, where the patched program cannot be compiled; \textbf{(2) runtime failures}, where the patched program compiles successfully but triggers runtime exceptions or timeouts during testing; \textbf{(3) remaining failures}, where the originally failing test(s) are still not fully fixed; and \textbf{(4) regression failures}, where the original failure is resolved but previously passing tests become failing. 
After establishing this coarse-grained diagnosis direction, \textsc{PracRepair} further extracts three complementary forms of feedback to understand \emph{why} the patch fails. 
First, it collects \emph{validation diagnostics}, such as compiler errors, runtime exceptions, timeout messages, or updated failing tests, to describe the observed failure outcome. 
Second, it computes a \emph{code diff} between the generated patch and the original buggy function to identify which statements or conditions have been changed. 
Third, it computes a \emph{trace diff} between the original and patched executions. 
To compute the trace diff, \textsc{PracRepair} executes the same triggering tests on both versions, collects traces using the same instrumentation procedure, aligns trace records by executed statement and execution order, and extracts changed branch outcomes, added or removed statement executions, and divergent runtime values. 
The resulting feedback therefore explains not only whether the patch fails, but also how the patch changes the failing behavior. 

This extracted feedback is then fed back into \emph{Question-driven Failure Diagnosis} as optional diagnostic input, as shown in Figure~\ref{fig:approach}. 
Based on the original bug context, the accumulated QA history, and the new feedback, the LLM re-diagnoses the current patch failure. 
For instance, if the trace diff shows that a branch outcome changes but the failing value remains abnormal, the next diagnosis round can ask why the changed branch still does not restore the expected state. 
The resulting QA pairs are used to refine the repair hypothesis, which then guides the next round of patch generation and validation. 
Through this feedback-guided loop, \textsc{PracRepair} progressively improves candidate patches until a plausible fix is found or the refinement budget is exhausted.
\section{Experiment Design}
To evaluate our approach, we design experiments to answer the following research questions (RQs):

\begin{itemize}[leftmargin=12pt, noitemsep, topsep=2pt]
    \item \textbf{RQ1 (Repair Effectiveness):} How effective is \textsc{PracRepair} compared with existing APR tools under the standard perfect fault localization setting, and does it remain effective when exact fault locations are unavailable?
    \item \textbf{RQ2 (Repair Scenarios):} How well does \textsc{PracRepair} perform across different repair scenarios?
    \item \textbf{RQ3 (Ablation Study):} What are the individual contributions of each component of \textsc{PracRepair} to the overall improvement in repair effectiveness?
    \item \textbf{RQ4 (Generalizability Study):} How effectively does \textsc{PracRepair} generalize to unseen datasets when deployed with different underlying foundation models?
\end{itemize}

\subsection{Datasets}
Since our approach is implemented and evaluated in the Java APR setting, we use two Java bug-repair benchmarks, Defects4J~\cite{defects4j} and RWB (Real-World Bugs)~\cite{thinkRepair}. We therefore do not include datasets in other programming languages, such as SWE-Bench, in this study. 
For the Defects4J dataset, following prior studies~\cite{chatRepair, thinkRepair, ReInFix}, we split it into V1.2 (391 bugs after removing 4 deprecated ones) and V2.0 (438 new bugs). We also follow~\cite{chatRepair, thinkRepair, ReInFix} to categorize bugs into four repair scenarios: multi-function (MF), where a fix involves multiple functions; single-function (SF), where a fix is confined to one function; single-hunk (SH), where a fix modifies one contiguous code region; and single-line (SL), where a fix changes only one line. Note that SH $\subseteq$ SF and SL $\subseteq$ SH. Table~\ref{tab:d4j-statistics} shows the statistics. 
For the generalizability study, we use the recent benchmark RWB (Real-World Bugs) introduced by ThinkRepair~\cite{thinkRepair},  which consists of two versions. RWB V1.0 comprises bug-fixing commits after October 2021, while RWB V2.0 includes bug-fixing commits after March 2023, resulting in 44 and 29 single-function bugs, respectively.
For fault information, to eliminate potential bias introduced by different fault localization (FL) tools, we follow recent APR studies~\cite{chatRepair, thinkRepair, repairAgent, ReInFix} and use perfect fault localization as the default setting, where the repair system is provided with the exact buggy statement location(s). In addition, to examine whether the effectiveness of \textsc{PracRepair} depends on this idealized assumption, we further include a relaxed fault-localization setting in RQ1, where exact buggy statement locations are not provided. 

\subsection{Implementation}
For the base models, we use \texttt{gpt-3.5-turbo}~\cite{openai2024gpt3.5} and \texttt{gpt-4o}~\cite{openai2024gpt4} in our main experiments to maintain direct comparability with prior APR studies. Following~\cite{ChatGPTOutperforms}, we set the sampling temperature to 1.0. To further evaluate whether \textsc{PracRepair} remains effective with newer foundation models, we additionally study its generalizability in RQ4 (Section~\ref{sec:generalizability_study}) using \texttt{gpt-4}~\cite{openai_gpt4_0613_2023}, \texttt{Llama-3}~\cite{ollama_llama3_2024}, and \texttt{DeepSeek-v3}~\cite{deepseekcoder2023}. 
We set the maximum number of repair sessions to 3 per bug, where each session is independent and starts from the original bug context. 
Within each session, the diagnosis loop is allowed to run for at most 10 rounds, although in practice the average number of rounds is no more than 5, since the loop terminates once no further diagnostic questions are raised. 
The refinement loop is allowed to run for at most 3 rounds, since this setting achieves a better balance between repair effectiveness (cf. Section~\ref{sec:ablation_study}).  
All experiments were conducted on a workstation running Ubuntu 20.04, with a 16-core Intel Xeon processor, 192GB of RAM, and eight NVIDIA A800 GPUs. 

\begin{table}[t]
\centering
\small
\setlength{\tabcolsep}{2.5pt} 
\renewcommand{\arraystretch}{0.9}
\caption{Statistics of Studied Datasets.}
\label{tab:d4j-statistics}
\resizebox{0.45\textwidth}{!}{
\begin{tabular}{lccccc}
\toprule
\textbf{Dataset} 
& \textbf{\#Total Bugs} 
& \textbf{\#MF Bugs} 
& \textbf{\#SF Bugs} 
& \textbf{\#SH Bugs} 
& \textbf{\#SL Bugs} \\
\midrule
Defects4J~1.2   & 391 & 136 & 255 & 154 & 80 \\
Defects4J~2.0   & 438 & 210 & 228 & 159 & 78 \\
\midrule
\textbf{\#Sum}  & \textbf{909} & \textbf{346} & \textbf{563} & \textbf{390} & \textbf{235} \\
\bottomrule
\end{tabular}
}
\end{table}

\begin{table*}[t]
\centering
\setlength{\tabcolsep}{4pt}
\caption{Repair results (correct patches / plausible patches) of APR tools under perfect fault localization on Defects4J. }
\label{tab:apr-comparison-pf}
\scriptsize
\renewcommand{\arraystretch}{1.1}

\resizebox{1\textwidth}{!}{%
\begin{tabular}{lccccccc cccc c}
\toprule
\textbf{APR Tool}  
& \textbf{\textsc{PracRepair}$_{\text{GPT-4o}}$}
& \textbf{\textsc{PracRepair}$_{\text{GPT-3.5}}$}
& \textbf{ReInFix$_{\text{GPT-4o}}$}
& \textbf{ReInFix$_{\text{GPT-3.5}}$}
& \textbf{ChatRepair}
& \textbf{ThinkRepair}
& \textbf{RepairAgent}

& \textbf{AlphaRepair}
& \textbf{KNOD}
& \textbf{Tare}
& \textbf{SelfRepair}

& \textbf{TBar}
\\

\midrule
Chart              
& 19/22 & 18/20 & 18/20 & 16/17 & 15/-- & 11/-- & 11/14 & 9/--  & 10/11  & 11/-- & 7/-- & 11/-- \\
Closure            
& 40/52 & 34/40 & 40/50 & 30/37 & 37/-- & 31/-- & 25/25 & 23/-- & 23/29 & 25/-- & 17/-- & 16/-- \\
Lang               
& 44/49 & 36/39 & 33/47 & 26/33 & 21/-- & 19/-- & 17/17 & 13/-- & 11/13 & 14/-- & 10/-- & 13/-- \\
Math               
& 42/70 & 38/54 & 39/68 & 35/52 & 32/-- & 27/-- & 29/29 & 21/-- & 20/25 & 22/-- & 18/-- & 22/-- \\
Mockito            
& 10/11 & 9/9 & 10/11 & 8/9   & 6/--  & 6/--  & 6/6   & 5/--  & 5/5  & 2/--  & 3/-- & 3/-- \\
Time               
& 7/9 & 4/5 & 6/11  & 3/4   & 3/--  & 4/--  & 2/3   & 3/--  & 2/2  & 3/--  & 3/-- & 3/-- \\
\midrule
\#Total (D4J V1.2) 
& \textbf{162/213} & \textbf{139/167} & 146/207 & 118/152 & 114/-- & 98/-- & 90/94 & 74/109 & 71/85 & 77/-- & 58/-- & 68/95\\
\#Total (D4J V2.0) 
& \textbf{171/200} & \textbf{136/165} & 145/190 & 123/147 & 48/--  & 107/-- & 74/92 & 36/--  & 50/85 & --/-- & 42/-- & 8/-- \\
\midrule
\#Sum
& \textbf{333/413} & \textbf{275/332} & 291/397 & 241/299 & 162/--  & 205/-- & 164/186 & 110/109  & 121/170 & 77/-- & 100/-- & 76/95 \\

\bottomrule
\end{tabular}%
}

\vspace{2pt}
\begin{minipage}{\textwidth}
\footnotesize
\emph{Note:} ``--'' indicates that no result was reported in the original work.
\end{minipage}
\end{table*}

\subsection{Baselines}
\label{baselines}
In our comparative evaluation, we evaluate \textsc{PracRepair} against nine state-of-the-art baselines. These baselines include one traditional APR method, TBar~\cite{Tbar}; three learning-based APR methods, SelfAPR~\cite{selfAPR}, KNOD~\cite{KNOD}, and Tare~\cite{Tare}; and five recent LLM-based APR methods, including Codex~\cite{Xi2023Automated}, AlphaRepair~\cite{AlphaRepair}, ChatRepair~\cite{chatRepair}, ThinkRepair~\cite{thinkRepair}, RepairAgent~\cite{repairAgent}, and ReinFix~\cite{ReInFix}. Since our evaluation adopts the same benchmark split, fault-localization setting, and repair metrics as these studies, we follow common practice in the APR community~\cite{selfAPR, AlphaRepair, chatRepair, thinkRepair, ReInFix} and reuse the repair results reported in their original papers~\cite{Tbar, selfAPR, KNOD, Tare, AlphaRepair, chatRepair, thinkRepair, repairAgent, ReInFix} instead of directly running these APR tools. 

To conduct the ablation study and investigate the contribution of different components of \textsc{PracRepair}, we design the following variants by removing or replacing components of the framework.
\begin{itemize}[leftmargin=12pt, noitemsep, topsep=2pt]
\item 
\textbf{w/o SDC+QFD+FPR:} This variant is designed to evaluate the overall contribution of the proposed three-stage framework. Specifically, it removes all three stages and directly prompts the underlying LLM to generate a patch from the given bug information.

\item 
\textbf{ w/o SDC+QFD:} This variant is designed to investigate the contribution of \emph{Static-dynamic Context Construction} and \emph{Question-driven Failure Diagnosis}. Specifically, it retains only \emph{Feedback-guided Patch Refinement}, allowing the model to iteratively improve patches based on feedback.

\item 
\textbf{w/o SDC:} This variant evaluates the contribution of \emph{Static-dynamic Context Construction}. It preserves \emph{Question-driven Failure Diagnosis} and \emph{Feedback-guided Patch Refinement}, but removes the structured repair context built from static and dynamic evidence. 

\item 
\textbf{w/o DI:} This variant evaluates the contribution of dynamic execution information in \emph{Static-dynamic Context Construction}. Specifically, it removes dynamic traces and retains only static program context with failure information.

\item 
\textbf{\textsc{PracRepair}$_{\text{CoT}}$:} 
This variant evaluates the contribution of the proposed diagnosis strategy in \emph{Question-driven Failure Diagnosis}. 
It replaces question-driven diagnosis with Chain-of-Thought prompting, 
in which the model formulates a repair hypothesis from the buggy code, failure information, and execution traces, 
similar to ThinkRepair~\cite{thinkRepair}. 

\item 
\textbf{\textsc{PracRepair}$_{\text{ReAct}}$:} 
This variant is designed to investigate the contribution of question-driven diagnosis. 
Specifically, it allows the LLM to use function calls, 
but replaces the proposed diagnosis strategy with direct interleaving of reasoning and tool use, 
similar to ReInFix~\cite{ReInFix}.

\end{itemize}
\subsection{Metrics}
\label{sec:metricss}
Following prior work~\cite{chatRepair,thinkRepair,repairAgent, ReInFix}, we report two widely adopted metrics to evaluate repair effectiveness:
\begin{itemize}[leftmargin=12pt, noitemsep, topsep=2pt]
\item \textbf{Number of plausible patches:} The number of bugs for which at least one generated patch passes all  developer-written test cases~\cite{KNOD, cure}. A plausible patch satisfies the test oracle but is not necessarily semantically correct.
\item \textbf{Number of correct patches:} The number of bugs for which at least one generated patch is semantically correct. To determine correctness, we first check whether a generated patch matches the developer-provided fix; otherwise, we manually assess its semantic equivalence. A patch is considered correct if it passes either of these checks~\cite{selfAPR}.
\end{itemize}

\section{Evaluation}
\subsection{RQ1: Repair Effectiveness} 
We evaluate the repair effectiveness of \textsc{PracRepair} on Defects4J under both the standard perfect fault localization setting and a relaxed setting where exact buggy statement locations are unavailable. Under the standard setting, we first compare \textsc{PracRepair} with existing APR tools on repair results, and then further analyze its unique repair capability. Under the relaxed setting, we examine whether \textsc{PracRepair} remains effective without perfect fault localization.

\noindent 
\textbf{Effectiveness under Perfect Fault Localization.}
Following the standard setting used in prior APR studies, we instantiate \textsc{PracRepair} with two foundation models, GPT-3.5~\cite{openai2024gpt3.5} and GPT-4o~\cite{openai2024gpt4}, referred to as \textsc{PracRepair}$_{\text{GPT-3.5}}$ and \textsc{PracRepair}$_{\text{GPT-4o}}$, respectively. 
As shown in Table~\ref{tab:apr-comparison-pf}, \textsc{PracRepair}$_{\text{GPT-3.5}}$ generates plausible fixes for 332 bugs and correct fixes for 275 bugs. 
With GPT-4o, \textsc{PracRepair}$_{\text{GPT-4o}}$ further improves to 413 plausible fixes and 333 correct fixes. 
Since plausible patches pass all test cases but are not necessarily semantically correct, these results indicate that \textsc{PracRepair} not only satisfies the test oracle on a large number of bugs, but also achieves strong repair accuracy. 
In addition, \textsc{PracRepair} successfully fixes bugs across all Defects4J projects, including Chart, Closure, Lang, Math, Mockito, and Time, demonstrating its effectiveness across projects from different domains. 

Compared with prior work, \textsc{PracRepair} consistently outperforms the strongest baseline, ReInFix, on both Defects4J V1.2 and V2.0. 
On Defects4J V1.2, \textsc{PracRepair}$_{\text{GPT-4o}}$ improves over ReInFix$_{\text{GPT-4o}}$ by 16 correct fixes, while \textsc{PracRepair}$_{\text{GPT-3.5}}$ exceeds ReInFix$_{\text{GPT-3.5}}$ by 21 fixes. 
Similar gains are observed on the more challenging Defects4J V2.0 benchmark, where \textsc{PracRepair}$_{\text{GPT-4o}}$ and \textsc{PracRepair}$_{\text{GPT-3.5}}$ outperform their ReInFix counterparts by 26 and 13 bugs, respectively. 
In addition, under GPT-3.5, \textsc{PracRepair} also surpasses other recent LLM-based APR approaches, including ChatRepair, ThinkRepair, and RepairAgent. 

\begin{figure}[t]
    \centering
    \subfloat[vs. GPT-3.5-based result]{
        \includegraphics[width=0.46\columnwidth]{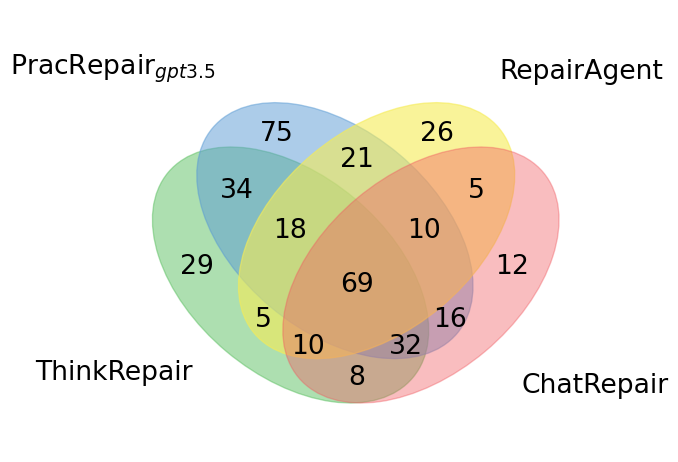}
        \label{fig:unique_fix_2}
    }
    \hfill
    \subfloat[vs. GPT-4o-based result]{
        \includegraphics[width=0.46\columnwidth]{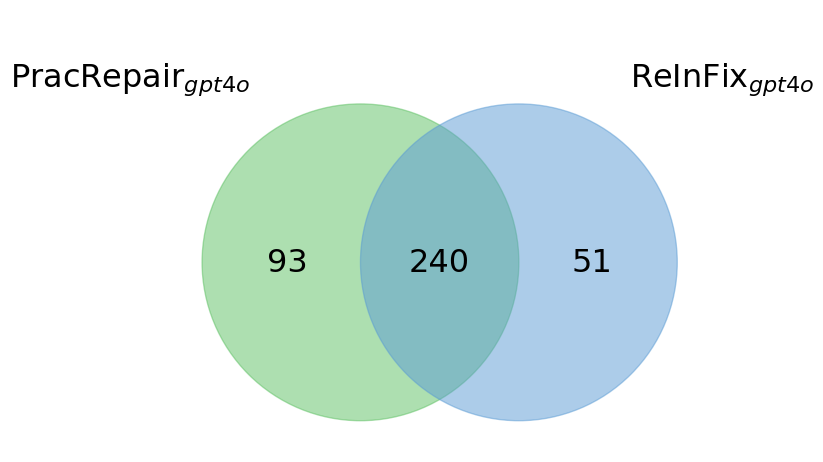}
        \label{fig:unique_fix_4}
    }

    \caption{Venn diagram of correct patches of \textsc{PracRepair} vs. LLM-based baselines on Defects4J V1.2 and V2.0.}
    \label{fig:unique_fix}
\end{figure}

\noindent
\textbf{Unique Fix Analysis.}
We further analyze the unique repair capability of \textsc{PracRepair} on Defects4J V1.2 and V2.0. Specifically, we compare the sets of correctly repaired bugs produced by \textsc{PracRepair} and recent LLM-based APR baselines under the same base model setting. As shown in Figure~\ref{fig:unique_fix}, under GPT-3.5, \textsc{PracRepair}$_{\text{GPT-3.5}}$ achieves 75 unique correct fixes, compared with 29 for ThinkRepair, 26 for RepairAgent, and 12 for ChatRepair. Under GPT-4o, \textsc{PracRepair}$_{\text{GPT-4o}}$ achieves 93 unique correct fixes, while ReInFix achieves 51. We exclude ReInFix from the GPT-3.5-based comparison because its public results are only available under GPT-4o. These results show that \textsc{PracRepair} maintains stronger unique repair capability than existing LLM-based APR baselines, suggesting that its repair process complements prior methods. 

\begin{table}[t]
\centering
\setlength{\tabcolsep}{2pt}
\caption{Repair results (correct patches / plausible patches) without perfect fault localization on Defects4J V1.2.}
\label{tab:apr-comparison-mf}
\resizebox{\columnwidth}{!}{
\begin{tabular}{lcccc}
\toprule

\textbf{APR Tool}  
& \textbf{\textsc{PracRepair}$_{\text{No-PFL}}$}
& \textbf{ThinkRepair$_{\text{No-PFL}}$}
& \textbf{Codex}
\\

\midrule
Chart              
& 12/16 & 9/--  & --/-- \\
Closure            
& 28/32 & 19/-- & --/-- \\
Lang               
& 23/31 & 15/-- & --/-- \\
Math               
& 32/43 & 27/-- & --/-- \\
Mockito            
& 7/8   & 7/--  & --/-- \\ 
Time               
& 3/3   & 3/--  & --/--\\
\midrule
\#Total (D4J V1.2)
& \textbf{105}/\textbf{133} & 80/--  & 63/-- \\
\bottomrule
\end{tabular}
}

\vspace{2pt}
\begin{minipage}{\columnwidth}
\footnotesize
\emph{Note:} ``--'' indicates that no result was reported in the original work.
\end{minipage}
\end{table}

\noindent \textbf{Effectiveness without Perfect Fault Localization.} 
The above comparisons assume perfect fault localization, where exact buggy statement locations are provided. To examine whether \textsc{PracRepair} remains effective without this assumption, we further evaluate it under GPT-3.5 without providing exact fault locations, referred to as \textsc{PracRepair}$_{\text{No-PFL}}$. We compare it with available baselines under the same setting, including ThinkRepair$_{\text{No-PFL}}$~\cite{thinkRepair} and Codex~\cite{Xi2023Automated}, on Defects4J V1.2. As shown in Table~\ref{tab:apr-comparison-mf}, \textsc{PracRepair}$_{\text{No-PFL}}$ fixes 105 bugs correctly and generates 133 plausible patches. 
Although this is lower than \textsc{PracRepair}$_{\text{GPT-3.5}}$ under perfect fault localization (139 correct and 167 plausible patches), it outperforms ThinkRepair$_{\text{No-PFL}}$ and Codex by 25 and 42 correct fixes, respectively. These results show that fault locations are helpful, but \textsc{PracRepair} remains effective when they are unavailable.

\begin{tcolorbox}[colback=gray!8, colframe=gray!50, arc=3mm,
    boxrule=0.5pt,
    left=1pt,
    right=1pt,
    top=1pt,
    bottom=1pt,
]
\textbf{Answer to RQ1:} \textsc{PracRepair} achieves the best overall repair effectiveness under perfect fault localization with both GPT-3.5 and GPT-4o. Without exact buggy statement locations, its performance decreases but still surpasses the only comparable baseline, showing that its effectiveness does not solely rely on perfect fault localization.
\end{tcolorbox}

\subsection{RQ2: Repair Scenarios}
\begin{table}[t]
\centering
\caption{Repair results (correct fixes) under different repair scenarios.}
\label{tab:scenario-results}
\scriptsize
\renewcommand{\arraystretch}{1.1}

\resizebox{\columnwidth}{!}{%
\begin{tabular}{l cccc cccc}
\toprule
\textbf{Benchmark}
& \multicolumn{4}{c}{\textbf{Defects4J V1.2}} 
& \multicolumn{4}{c}{\textbf{Defects4J V2.0}} \\
\cmidrule(lr){2-5}
\cmidrule(lr){6-9}

\textbf{Repair Scenario} & MF & SF & SH & SL & MF & SF & SH & SL \\
\midrule
ChatRepair                  & -- & 76  & -- & -- & -- & --  & -- & 48 \\
ThinkRepair                 & -- & 98  & 78 & 52 & -- & 107 & 81 & 47 \\
RepairAgent                 & 7  & 83  & 71 & 51 & 6  & 68  & 65 & 48 \\
ReInFix$_{\text{GPT-3.5}}$  & 14 & 104 & 78 & 53 & 14 & 109 & 85 & 47 \\
ReInFix$_{\text{GPT-4o}}$   & 22 & 124 & 93 & 57 & 15 & 130 & 103 & 56 \\
\textsc{PracRepair}$_{\text{GPT-3.5}}$      & \textbf{19} & \textbf{120}  & \textbf{90} & \textbf{55} & \textbf{15} & \textbf{121}  & \textbf{92} & \textbf{51} \\
\textsc{PracRepair}$_{\text{GPT-4o}}$       & \textbf{27} & \textbf{135}  & \textbf{97} & \textbf{57} & \textbf{18} & \textbf{153}  & \textbf{108} & \textbf{57} \\

\bottomrule
\end{tabular}
}

\vspace{2pt}
\begin{minipage}{\columnwidth}
\footnotesize
\emph{Note:} ``--'' indicates that no result was reported in the original work.
\end{minipage}
\end{table}

While RQ1 evaluates the overall repair effectiveness of \textsc{PracRepair}, analyzing its performance under specific repair scenarios provides a more fine-grained understanding of its capabilities across different levels of repair complexity. 
Following prior work~\cite{ReInFix, thinkRepair, repairAgent}, we further examine \textsc{PracRepair} under four commonly scenarios: single-line (SL), single-hunk (SH), single-function (SF), and multi-function (MF).

\noindent
\textbf{Repair Scenarios Analysis.}
As shown in Table~\ref{tab:scenario-results}, \textsc{PracRepair} achieves strong performance across all studied repair scenarios on both Defects4J V1.2 and V2.0. In the single-function (SF) setting, it delivers the best overall results among all compared approaches. On Defects4J V1.2, \textsc{PracRepair}$_{\text{GPT-3.5}}$ and \textsc{PracRepair}$_{\text{GPT-4o}}$ repair 120 and 135 bugs, respectively, while on Defects4J V2.0 the corresponding numbers further increase to 121 and 153, consistently exceeding all baselines.
In the single-hunk (SH) and single-line (SL) settings, \textsc{PracRepair} continues to match or surpass recent LLM-based baselines, demonstrating its effectiveness across simpler and more complex repair scenarios. More importantly, \textsc{PracRepair} shows clear advantages in the multi-function (MF) setting. Compared with ReInFix, the only other baseline explicitly supporting MF repair, \textsc{PracRepair} achieves higher repair counts on both dataset versions. For example, \textsc{PracRepair}$_{\text{GPT-4o}}$ repairs 27 and 18 MF bugs on Defects4J V1.2 and V2.0, compared with 22 and 15 for ReInFix$_{\text{GPT-4o}}$. Overall, these results indicate that \textsc{PracRepair} performs robustly across diverse repair scenarios, with particularly strong advantages on challenging multi-function bugs.

\begin{tcolorbox}[colback=gray!8, colframe=gray!50, arc=3mm,
    boxrule=0.5pt,
    left=1pt,
    right=1pt,
    top=1pt,
    bottom=1pt,
    width=\columnwidth,
]
\textbf{Answer to RQ2:} \textsc{PracRepair} consistently outperforms prior methods across different repair scenarios, including SL, SH, SF, and MF bugs. Its advantage is especially clear on the more challenging multi-function setting.
\end{tcolorbox}

\subsection{RQ3: Ablation Study}
\label{sec:ablation_study}
To assess the impact of individual components in \textsc{PracRepair}, we conduct an ablation study using the variants defined in Section~\ref{baselines}. 
These variants are designed by systematically removing or replacing key parts of the framework. 
Based on this design, we evaluate the contribution of each stage, as well as the effects of dynamic execution traces, diagnosis strategy, and refinement rounds. 
Due to computational budget constraints, all ablation experiments are conducted on Defects4J V1.2 with GPT-3.5.

\noindent
\textbf{Impacts of the Three Stages.}
Table~\ref{tab:ablation_core} reports the performance of several ablated variants, each designed to isolate the contribution of one stage in \textsc{PracRepair}. The \emph{w/o SDC+QFD+FPR} variant performs the worst, achieving 84 correct patches and 98 plausible patches. Adding only \emph{Feedback-guided Patch Refinement} in \emph{w/o SDC+QFD} increases the number of correct patches to 105, showing the benefit of iterative refinement. Further adding \emph{Question-driven Failure Diagnosis} in \emph{w/o SDC} raises the number of correct patches to 115, indicating that diagnosis improves repair beyond refinement alone. The full \textsc{PracRepair} configuration achieves the best results. Compared with \emph{w/o SDC}, adding \emph{Static-dynamic Context Construction} increases the number of correct patches from 115 to 139 and plausible patches from 126 to 167. Overall, the results show that all three stages contribute to repair effectiveness, and their combination yields the strongest performance. 

\begin{table}[t]
\centering
\renewcommand{\arraystretch}{1.0}
\caption{Repair results (correct patches / plausible patches) of the ablation study on Defects4J V1.2.}
\label{tab:ablation_core}

\begin{tabular}{lc|cccc} %
\toprule
\textbf{Variant}
& \textbf{Result} 
& \textbf{MF} 
& \textbf{SF} 
& \textbf{SH} 
& \textbf{SL} 
\\
\midrule
w/o SDC+QFD+FPR   & 84/98   & 7    & 77   & 55   & 38  \\
w/o SDC+QFD       & 105/113  & 12   & 93   & 74   & 46  \\
w/o SDC           & 115/126  & 16   & 99   & 77   & 48  \\
w/o DI            & 120/137  & 17   & 103  & 80   & 49  \\
\textsc{PracRepair}$_{\text{COT}}$      & 107/123  & 9    & 98   & 68   & 52  \\
\textsc{PracRepair}$_{\text{ReAct}}$    & 121/149  & 15   & 106  & 79   & 53  \\
\textsc{PracRepair}$_{\text{GPT-3.5}}$  & \textbf{139}/\textbf{167}  & \textbf{19}   & \textbf{120}  & \textbf{90}   & \textbf{55}  \\
\bottomrule
\end{tabular}

\vspace{2pt}
\begin{minipage}{\columnwidth}
\footnotesize
\emph{Note:} ``--'' indicates that no result was reported in the original work.
\end{minipage}

\end{table}

\begin{figure}[t]
    \centering
    \includegraphics[width=0.45\textwidth]{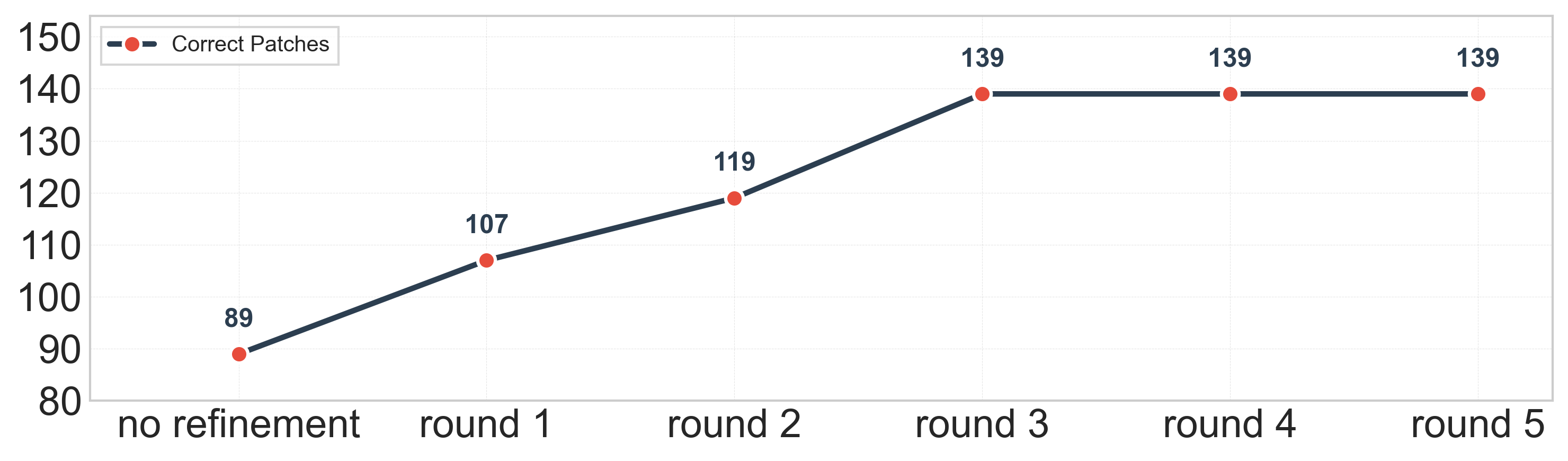}
    \caption{The performance of \textsc{PracRepair} with different refinement rounds}
    \label{fig:refine_round}
\end{figure}

\noindent
\textbf{Impacts of Dynamic Execution Traces.}
Table~\ref{tab:ablation_core} shows the contribution of dynamic execution traces to \textsc{PracRepair}. 
The \emph{w/o DI} variant removes dynamic execution traces from \emph{Static-dynamic Context Construction}, leaving only static program context and failure information. 
Compared with the full system, this change reduces the number of correct patches from 139 to 120 and the number of plausible patches from 167 to 137. 
These results indicate that dynamic execution traces provide important failure-relevant evidence that cannot be fully recovered from static context alone. 
By exposing runtime behaviors, such as executed paths, branch outcomes, and variable state changes, they help the model perform more grounded diagnoses and repairs. 

\noindent
\textbf{Impacts of Reasoning Strategy.}
Table~\ref{tab:ablation_core} shows that, compared with plain CoT and ReAct, the proposed question-driven diagnosis mechanism in \emph{Question-driven Failure Diagnosis} improves repair effectiveness. The one-shot \textsc{PracRepair}${\text{CoT}}$ variant, which does not use the designed function calls, produces 107 correct patches. Allowing on-demand retrieval of failure-relevant evidence in \textsc{PracRepair}${\text{ReAct}}$ increases this number to 121, which \emph{suggests} that tool-assisted diagnosis can be more effective than reasoning over a fixed input context alone. The full \textsc{PracRepair} further improves the result to 139. Since both \textsc{PracRepair}$_{\text{ReAct}}$ and the full \textsc{PracRepair} support function-call interaction, this additional gain \emph{indicates} that the proposed question-driven diagnosis provides benefits beyond tool use alone. By organizing diagnosis around targeted questions, \textsc{PracRepair} appears to help the model systematically inspect failure-relevant evidence and formulate repair hypotheses. 

\noindent\textbf{Impacts of Refinement Interaction Number.}
According to Figure~\ref{fig:refine_round}, repair performance improves as the number of refinement interactions increases. 
Without refinement, \textsc{PracRepair} produces 89 correct patches, which increases to 107 and 119 after one and two refinement rounds, respectively, indicating that iterative feedback helps correct early patch errors. 
Performance further improves to 139 correct patches after three rounds, but remains unchanged with additional interactions, showing diminishing returns beyond this point. 
Considering both repair effectiveness and interaction cost, we adopt three refinement rounds as the default setting.

\enlargethispage{\baselineskip}
\begin{tcolorbox}[colback=gray!8, colframe=gray!50, arc=3mm,
    boxrule=0.5pt,
    left=1pt,
    right=1pt,
    top=1pt,
    bottom=1pt,
    width=\columnwidth,
]
\textbf{Answer to RQ3:} \textsc{PracRepair} is well designed, and all three stages, i.e., Static-dynamic Context Construction, Question-driven Failure Diagnosis, and Feedback-guided Patch Refinement, can be effectively integrated to improve the correct repair effectiveness of \textsc{PracRepair}.
\end{tcolorbox}

\subsection{RQ4: Generalizability Study}
\label{sec:generalizability_study}
In the generalizability study, we evaluate \textsc{PracRepair} on the RWB benchmark under the perfect fault localization setting, following ThinkRepair~\cite{thinkRepair}, and further instantiate \textsc{PracRepair} with multiple foundation models, including GPT-4, GPT-3.5, DeepSeek-v3, DeepSeek-Coder, and Llama-3. We also report the published results of ThinkRepair~\cite{thinkRepair} and ReInFix~\cite{ReInFix} on the same benchmark for comparison. Notably, RWB V1.0 and RWB V2.0 consist of bug-fixing commits collected after the training cutoff dates of GPT-3.5 and DeepSeek-Coder, respectively~\cite{thinkRepair}.

\begin{table}[t]
\centering
\caption{Repair results (correct fixes) of the generalizability study on RWB.}
\label{tab:Generalizability}
\scriptsize
\setlength{\tabcolsep}{2.0pt}
\renewcommand{\arraystretch}{1.05}

\resizebox{\columnwidth}{!}{%
\begin{tabular}{lccccc ccc cc}
\toprule
\textbf{APR Tool}
& \multicolumn{5}{c}{\textbf{\textsc{PracRepair}}}
& \multicolumn{3}{c}{\textbf{ReInFix}}
& \multicolumn{2}{c}{\textbf{ThinkRepair}} \\
\cmidrule(lr){2-6}
\cmidrule(lr){7-9}
\cmidrule(l){10-11}

\textbf{LLM}
& G4 & DS-v3 & L3 & G3.5 & DS-C
& G4 & G3.5 & DS-C
& G3.5 & DS-C \\
\midrule

Cli         & 4 & 4 & 4 & 4 & -- & 4 & 4 & -- & 4 & -- \\
Codec       & 4 & 3 & 3 & 3 & -- & 3 & 3 & -- & 3 & -- \\
Collections & 1 & 1 & 1 & 1 & -- & 1 & 1 & -- & 1 & -- \\
Compress    & 3 & 3 & 3 & 2 & -- & 2 & 2 & -- & 1 & -- \\
Csv         & 1 & 1 & 1 & 1 & -- & 1 & 1 & -- & 1 & -- \\
Jsoup       & 7 & 8 & 7 & 7 & -- & 7 & 6 & -- & 6 & -- \\
Lang        & 3 & 3 & 3 & 3 & -- & 3 & 3 & -- & 3 & -- \\
\midrule

\# RWB V1.0 & \textbf{23} & \textbf{23} & 22 & 21 & -- & 21 & 20 & -- & 19 & -- \\
\# RWB V2.0 & \textbf{15} & 14 & 13 & -- & 13 & -- & -- & 12 & -- & 10 \\

\bottomrule
\end{tabular}%
}

\vspace{2pt}
\begin{minipage}{\columnwidth}
\footnotesize
\emph{Note:} ``--'' indicates that no result was reported in the original work.
G4/G3.5 = GPT-4/GPT-3.5; DS-v3/DS-C = DeepSeek-v3/DeepSeek-Coder; L3=Llama-3.
\end{minipage}
\end{table}

\noindent
\textbf{Result Analysis.}
As shown in Table~\ref{tab:Generalizability}, \textsc{PracRepair} achieves the best or tied-best repair performance across both RWB datasets and all evaluated model settings. On RWB V1.0 (44 bugs), \textsc{PracRepair} repairs 23 bugs with GPT-4 and DeepSeek-v3, outperforming ReInFix (21 bugs with GPT-4) and ThinkRepair (19 bugs with GPT-3.5). Similar trends hold under other models: \textsc{PracRepair} repairs 22 bugs with Llama-3 and 21 bugs with GPT-3.5, indicating stable effectiveness across different model backbones.
On the more recent RWB V2.0 benchmark (29 bugs), \textsc{PracRepair} again achieves the strongest results, repairing 13 bugs with DeepSeek-Coder, compared with 12 and 10 repaired by ReInFix and ThinkRepair, respectively. Moreover, when instantiated with open-source models such as GPT-4, DeepSeek-v3, and Llama-3, \textsc{PracRepair} still maintains competitive repair effectiveness. These results suggest that the proposed approach generalizes well across both benchmarks and foundation models, rather than depending on a specific dataset or model family.

\begin{tcolorbox}[colback=gray!8, colframe=gray!50, arc=3mm,
    boxrule=0.5pt,
    left=1pt,
    right=1pt,
    top=1pt,
    bottom=1pt,
    width=\columnwidth
]
\textbf{Answer to RQ4:} \textsc{PracRepair} generalizes well across both unseen benchmarks and different foundation models. These results show that its repair framework is robust and not tied to a specific dataset or model.
\end{tcolorbox}

\section{Discussion}
\subsection{Repair Costs}
Using LLMs may raise concerns about repair costs. To address this, we follow prior work~\cite{chatRepair, repairAgent, ReInFix} and report the average monetary cost per repaired bug based on the RQ1 results on Defects4J. For prior methods, we compare against the costs reported in their original papers. In our setting, \textsc{PracRepair}$_{\text{GPT-3.5}}$ costs \$0.04 per repaired bug, while \textsc{PracRepair}$_{\text{GPT-4o}}$ costs \$1.13. Compared with prior LLM-based APR methods, \textsc{PracRepair} remains cost-efficient: under GPT-3.5, its cost is lower than ReInFix$_{\text{GPT-3.5}}$ (\$0.06), RepairAgent (\$0.14), and ChatRepair (\$0.42), while under GPT-4o it also costs less than ReInFix$_{\text{GPT-4o}}$ (\$1.45). These results show that \textsc{PracRepair} improves repair effectiveness while maintaining competitive repair cost.

\subsection{Threats to Validity}

\noindent \textbf{Internal Validity.}
One internal threat comes from the manual validation of plausible patches. Since passing all test cases does not guarantee semantic correctness, we first check whether a plausible patch exactly matches the developer-provided fix; otherwise, we manually assess its semantic equivalence, following prior APR work. 
Another threat comes from potential data leakage, as some benchmark bugs or reference patches may have appeared in the pre-training data of the evaluated LLMs. To mitigate this concern, we additionally evaluate \textsc{PracRepair} on the RWB benchmark, whose bug-fixing commits were collected after the training cutoff dates of widely used LLMs. \textsc{PracRepair} still achieves strong results on this benchmark under multiple foundation models, suggesting that its gains are not merely due to memorization. 

\noindent \textbf{External Validity.} To reduce the risk of an unrepresentative evaluation, we assess \textsc{PracRepair} on Defects4J and RWB, two widely used real-world Java bug benchmarks. 
However, both datasets are limited to Java and may not fully represent other programming languages or much larger codebases. Evaluating \textsc{PracRepair} on additional languages and broader repair settings remains future work.

\section{Related Work}
Existing APR approaches can be broadly categorized from three perspectives: non-learning-based approaches, learning-based approaches, and LLM-based approaches.

\noindent
\textbf{Non-learning-based Approaches.}
Automated program repair has been widely studied for more than a decade~\cite{le2019automated}. Early approaches formulate repair as a search problem with manually designed mutation operators or fix patterns~\cite{le2012genProg, Tbar}. Other techniques learn transformation templates or repair patterns from human-written patches~\cite{Kim2013Automatic, Bavishi2019Phoenix}, or synthesize repairs using symbolic execution, constraints, and SMT solving~\cite{Nguyen2013SemFix, xiao2013characteristic}. Additional work integrates repair into static analysis pipelines or retrieves similar code fragments as repair ingredients~\cite{liu2023program, Marginean2019SapFix}. Beyond functional bugs, prior studies have also addressed syntax errors, performance bugs, vulnerabilities, type errors, and build failures~\cite{Gupta2019Deep, Tarlow2020Learning}.

\noindent
\textbf{Learning-based Approaches.}
With the development of machine learning, repair methods increasingly rely on data-driven models. Early learning-based methods use machine learning to rank or prioritize candidate patches~\cite{Long2016Automatic}. More recent approaches adopt neural machine translation models to directly transform buggy code into fixed code~\cite{Gupta2017DeepFix, Tufano2019On}, or design neural architectures that predict tree-level or syntax-aware code transformations~\cite{zhu2021syntax, Li2020DLFix}. Some methods further train repair-specific models on curated bug-fix datasets~\cite{Ye2022Neural, selfAPR}. Unlike these task-specific learning approaches, recent LLM-based APR methods use general-purpose foundation models without explicit repair-specific training.

\noindent
\textbf{LLM-based Approaches.}
With the emergence of large language models, APR has increasingly shifted toward prompt-based and agentic repair paradigms. Early LLM-based approaches mainly rely on prompt engineering to perform one-shot repair, where the model directly generates a candidate patch from buggy code and related inputs in a single interaction~\cite{Xi2023Automated, Jiang2023Impact, AlphaRepair}. Later methods introduce iterative repair by repeatedly querying the LLM with validation feedback and refining patches across multiple rounds~\cite{chatRepair, thinkRepair, chatdbg}. More recent agent-based approaches further extend this paradigm by allowing the LLM to invoke external tools during repair~\cite{repairAgent, ReInFix}. Our work is most closely related to this line of research, but differs in that it is inspired by practical debugging behaviors and structures repair around static-dynamic information integration, question-driven failure diagnosis, and feedback-guided patch refinement. 
\section{Conclusion}
In this work, we present \textsc{PracRepair}, a fully automated program repair framework inspired by real-world debugging practices. Specifically, \textsc{PracRepair} constructs static and dynamic context, performs question-driven failure diagnosis to formulate explicit repair hypotheses, and iteratively refines candidate patches using validation feedback. 
Extensive experiments, including comparisons with state-of-the-art baselines, scenario-based analysis, and ablation studies, show that \textsc{PracRepair} consistently outperforms existing methods. These results suggest that developer-inspired debugging workflows can substantially improve APR effectiveness. 
In future work, we plan to extend \textsc{PracRepair} to more languages for stronger generalizability.

\bibliography{reference}

\end{sloppypar}
\end{document}